\pdfoutput=1
\documentclass[12pt,a4paper]{article}
\usepackage{booktabs}
\usepackage{ifthen} % for conditional statements
\newboolean{pdflatex}
\setboolean{pdflatex}{true} % False for eps figures 
\usepackage{siunitx}
\newboolean{articletitles}
\setboolean{articletitles}{true} % False removes titles in references

\newboolean{uprightparticles}
\setboolean{uprightparticles}{false} %True for upright particle symbols

\def\paperauthors{LHCb collaboration} % Leave as is for PAPER, CONF and FIGURE
\def\paperasciititle{Measurement of CP violation observables in D+ -> K-K+pi+ decays} % Set ASCII title here !! MAKE sure it's only ASCII characters !! 
\def\papertitle{Measurement of  \\{{\ensuremath{C\!P}}\xspace} violation observables \\ in {\ensuremath{\D^+\rightarrow K^-K^+\pi^+}\xspace} decays} % Latex formatted title
\def\paperkeywords{{High Energy Physics}, {LHCb}} % Comma separated list
\def\papercopyright{\the\year\ CERN for the benefit of the LHCb collaboration} % new since 9/Apr/2018
\def\paperlicence{CC BY 4.0 licence}
\def\paperlicenceurl{https://creativecommons.org/licenses/by/4.0/}

\newif\ifEnableSectionTOCLinks
\EnableSectionTOCLinksfalse % deactivated

\usepackage[top=1in, bottom=1.25in, left=1in, right=1in]{geometry}

\usepackage{microtype}
\usepackage{lineno}  % for line numbering during review
\usepackage{xspace} % To avoid problems with missing or double spaces after
\usepackage{caption} %these three command get the figure and table captions automatically small

\usepackage{graphicx}  % to include figures (can also use other packages)
\usepackage{color}
\usepackage{colortbl}
\graphicspath{{./figs/}} % Make Latex search fig subdir for figures
\usepackage{amsmath} % Adds a large collection of math symbols
\usepackage{amssymb}
\usepackage{amsfonts}
\usepackage{upgreek} % Adds in support for greek letters in roman typeset

\newcommand*\patchAmsMathEnvironmentForLineno[1]{%
\expandafter\let\csname old#1\expandafter\endcsname\csname #1\endcsname
\expandafter\let\csname oldend#1\expandafter\endcsname\csname
end#1\endcsname
 \renewenvironment{#1}%
   {\linenomath\csname old#1\endcsname}%
   {\csname oldend#1\endcsname\endlinenomath}%
}
\newcommand*\patchBothAmsMathEnvironmentsForLineno[1]{%
  \patchAmsMathEnvironmentForLineno{#1}%
  \patchAmsMathEnvironmentForLineno{#1*}%
}
\AtBeginDocument{%
\patchBothAmsMathEnvironmentsForLineno{equation}%
\patchBothAmsMathEnvironmentsForLineno{align}%
\patchBothAmsMathEnvironmentsForLineno{flalign}%
\patchBothAmsMathEnvironmentsForLineno{alignat}%
\patchBothAmsMathEnvironmentsForLineno{gather}%
\patchBothAmsMathEnvironmentsForLineno{multline}%
\patchBothAmsMathEnvironmentsForLineno{eqnarray}%
}

\usepackage{hyperxmp}

\usepackage[pdftex,
            pdfauthor={\paperauthors},
            pdftitle={\paperasciititle},
            pdfkeywords={\paperkeywords},
            pdfcopyright={Copyright (C) \papercopyright},
            pdflicenseurl={\paperlicenceurl}]{hyperref}
\usepackage[colorinlistoftodos,textsize=scriptsize]{todonotes}

\usepackage[bottom,flushmargin,hang,multiple]{footmisc}

\usepackage[all]{hypcap} % Internal hyperlinks to floats.

\usepackage{xspace} 
\usepackage{upgreek}

\def\lhcb   {\mbox{LHCb}\xspace}

\def\babar  {\mbox{BaBar}\xspace}

\def\cleo   {\mbox{CLEO}\xspace}

\def\MagUp {\mbox{\em Mag\kern -0.05em Up}\xspace}

\ifthenelse{\boolean{uprightparticles}}%
{

 \def\Ppi         {\ensuremath{\uppi}\xspace}

 \def\Pphi        {\ensuremath{\upphi}\xspace}

 \def\PDelta      {\ensuremath{\Delta}\xspace}                 
 \def\PXi         {\ensuremath{\Xi}\xspace}                 
 \def\PLambda     {\ensuremath{\Lambda}\xspace}                 
 \def\PSigma      {\ensuremath{\Sigma}\xspace}                 
 \def\POmega      {\ensuremath{\Omega}\xspace}                 
 \def\PUpsilon    {\ensuremath{\Upsilon}\xspace}
 \let\oldPi\Pi
 \def\PPi         {\ensuremath{\oldPi}\xspace}

 \def\PB      {\ensuremath{\mathrm{B}}\xspace}                 
 \def\PD      {\ensuremath{\mathrm{D}}\xspace}                 

 \def\PK      {\ensuremath{\mathrm{K}}\xspace}                 
 \def\Pb      {\ensuremath{\mathrm{b}}\xspace}                 
 \def\Pc      {\ensuremath{\mathrm{c}}\xspace}

 \def\Pp      {\ensuremath{\mathrm{p}}\xspace}                 

 \def\Ps      {\ensuremath{\mathrm{s}}\xspace}

 \def\thebaroffset{0.0em}
}
{

 \def\Ppi         {\ensuremath{\pi}\xspace}

 \def\Pphi        {\ensuremath{\phi}\xspace}

 \mathchardef\PDelta="7101
 \mathchardef\PXi="7104
 \mathchardef\PLambda="7103
 \mathchardef\PSigma="7106
 \mathchardef\POmega="710A
 \mathchardef\PUpsilon="7107
 \mathchardef\PPi="7105
 \def\PB      {\ensuremath{B}\xspace}                 
 \def\PD      {\ensuremath{D}\xspace}                 

 \def\PK      {\ensuremath{K}\xspace}                 
 \def\Pb      {\ensuremath{b}\xspace}                 
 \def\Pc      {\ensuremath{c}\xspace}

 \def\Pp      {\ensuremath{p}\xspace}                 

 \def\Ps      {\ensuremath{s}\xspace}

 \def\thebaroffset{0.18em}
}
\newcommand{\offsetoverline}[2][\thebaroffset]{\kern #1\overline{\kern -#1 #2}}%

\makeatletter
\ifcase \@ptsize \relax% 10pt
  \newcommand{\miniscule}{\@setfontsize\miniscule{4}{5}}% \tiny: 5/6
\or% 11pt
  \newcommand{\miniscule}{\@setfontsize\miniscule{5}{6}}% \tiny: 6/7
\or% 12pt
  \newcommand{\miniscule}{\@setfontsize\miniscule{5}{6}}% \tiny: 6/7
\fi
\makeatother

\DeclareRobustCommand{\optbar}[1]{\shortstack{{\miniscule (\rule[.5ex]{1.25em}{.18mm})}
  \\ [-.7ex] $#1$}}

\def\squark    {{\ensuremath{\Ps}}\xspace}

\def\cquark    {{\ensuremath{\Pc}}\xspace}

\def\bquark    {{\ensuremath{\Pb}}\xspace}

\def\pion   {{\ensuremath{\Ppi}}\xspace}

\def\pip    {{\ensuremath{\pion^+}}\xspace}
\def\pim    {{\ensuremath{\pion^-}}\xspace}

\def\kaon    {{\ensuremath{\PK}}\xspace}
\def\Kbar    {{\ensuremath{\offsetoverline{\PK}}}\xspace}

\def\KorKbar {\kern \thebaroffset\optbar{\kern -\thebaroffset \PK}{}\xspace}

\def\Kp      {{\ensuremath{\kaon^+}}\xspace}
\def\Km      {{\ensuremath{\kaon^-}}\xspace}

\def\Kstarz  {{\ensuremath{\kaon^{*0}}}\xspace}
\def\Kstarzb {{\ensuremath{\Kbar{}^{*0}}}\xspace}

\newcommand{\phiz}{\ensuremath{\Pphi}\xspace}

\def\D       {{\ensuremath{\PD}}\xspace}

\def\DorDbar {\kern \thebaroffset\optbar{\kern -\thebaroffset \PD}\xspace}
\def\Dz      {{\ensuremath{\D^0}}\xspace}

\def\Dp      {{\ensuremath{\D^+}}\xspace}
\def\Dm      {{\ensuremath{\D^-}}\xspace}

\def\DpDm    {\ensuremath{\Dp {\kern -0.16em \Dm}}\xspace}

\def\Dsp     {{\ensuremath{\D^+_\squark}}\xspace}

\def\DporDsp {{\ensuremath{\D_{(\squark)}^+}}\xspace}
\def\DmorDsm {{\ensuremath{\D{}_{(\squark)}^-}}\xspace}
\def\DpmorDspm {{\ensuremath{\D{}_{(\squark)}^\pm}}\xspace}

\def\B       {{\ensuremath{\PB}}\xspace}

\def\BorBbar {\kern \thebaroffset\optbar{\kern -\thebaroffset \PB}\xspace}

\def\Bd      {{\ensuremath{\B^0}}\xspace}

\def\BdorBdbar {\kern \thebaroffset\optbar{\kern -\thebaroffset \Bd}\xspace}
\def\Bu      {{\ensuremath{\B^+}}\xspace}

\def\Bp      {{\ensuremath{\Bu}}\xspace}

\def\Bs      {{\ensuremath{\B^0_\squark}}\xspace}

\def\BsorBsbar {\kern \thebaroffset\optbar{\kern -\thebaroffset \Bs}\xspace}

\def\Y#1S{\ensuremath{\PUpsilon{(#1S)}}\xspace}

\def\proton      {{\ensuremath{\Pp}}\xspace}

\def\Lz          {{\ensuremath{\PLambda}}\xspace}

\def\LorLbar     {\kern \thebaroffset\optbar{\kern -\thebaroffset \PLambda}\xspace}

\def\Lc          {{\ensuremath{\Lz^+_\cquark}}\xspace}

\newcommand{\decay}[2]{\ensuremath{#1\!\to #2}\xspace} 

\def\to                 {\ensuremath{\rightarrow}\xspace}

\def\CP                {{\ensuremath{C\!P}}\xspace}

\def\AT#1     {\ensuremath{A_{\mathrm{T}}^{#1}}\xspace}           % 2

\def\C#1      {\ensuremath{\mathcal{C}_{#1}}\xspace}                       % 9
\def\Cp#1     {\ensuremath{\mathcal{C}_{#1}^{'}}\xspace}                    % 7
\def\Ceff#1   {\ensuremath{\mathcal{C}_{#1}^{\mathrm{(eff)}}}\xspace}        % 9  
\def\Cpeff#1  {\ensuremath{\mathcal{C}_{#1}^{'\mathrm{(eff)}}}\xspace}       % 7
\def\Ope#1    {\ensuremath{\mathcal{O}_{#1}}\xspace}                       % 2
\def\Opep#1   {\ensuremath{\mathcal{O}_{#1}^{'}}\xspace}                    % 7

\newcommand{\aunit}[1]{\ensuremath{\text{\,#1}}}       
\newcommand{\tev}{\aunit{Te\kern -0.1em V}\xspace}
\newcommand{\gev}{\aunit{Ge\kern -0.1em V}\xspace}
\newcommand{\mev}{\aunit{Me\kern -0.1em V}\xspace}
\newcommand{\kev}{\aunit{ke\kern -0.1em V}\xspace}
\newcommand{\ev}{\aunit{e\kern -0.1em V}\xspace}
 
\newcommand{\mevc}{\ensuremath{\aunit{Me\kern -0.1em V\!/}c}\xspace}
\newcommand{\gevc}{\ensuremath{\aunit{Ge\kern -0.1em V\!/}c}\xspace}
\newcommand{\mevcc}{\ensuremath{\aunit{Me\kern -0.1em V\!/}c^2}\xspace}
\newcommand{\gevcc}{\ensuremath{\aunit{Ge\kern -0.1em V\!/}c^2}\xspace}
\def\fb   {\ensuremath{\aunit{fb}}\xspace}
\def\invfb   {\ensuremath{\fb^{-1}}\xspace}

\newcommand{\chisq}{\ensuremath{\chi^2}\xspace}

\def\gsim{{~\raise.15em\hbox{$>$}\kern-.85em
          \lower.35em\hbox{$\sim$}~}\xspace}
\def\lsim{{~\raise.15em\hbox{$<$}\kern-.85em
          \lower.35em\hbox{$\sim$}~}\xspace}

\def\sPlot{\mbox{\em sPlot}\xspace}

\def\tell1  {TELL1\xspace}
\def\ukl1   {UKL1\xspace}

\newcommand{\ie}{\mbox{\itshape i.e.}\xspace}

\newcommand{\lhcborcid}[1]{\href{https://orcid.org/#1}{\hspace*{0.1em}\raisebox{-0.45ex}{\includegraphics[width=1em]{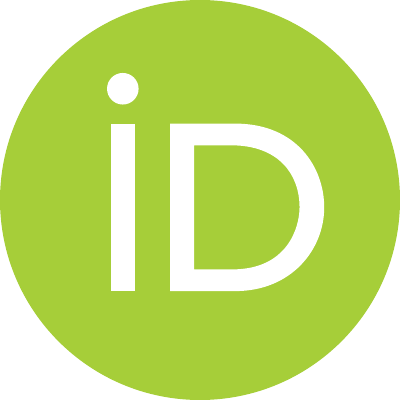}}}}

\hypersetup{
  colorlinks   = true, %Colours links instead of ugly boxes
  urlcolor     = blue, %Colour for external hyperlinks
  linkcolor    = blue, %Colour of internal links
  citecolor    = red   %Colour of citations
}

\ifEnableSectionTOCLinks
    \usepackage[explicit]{titlesec} % to change headings
    
    \let\oldcontentsline\contentsline
    \renewcommand

    \titleformat{\section}{\normalfont\Large\bf}{\hyperlink{tocsection.\thesection}{{\thesection} \parbox[t]{\dimexpr\textwidth-1pc}{#1}}}{1pc}{}

    \titleformat{\subsection}{\normalfont\bf}{\hyperlink{tocsubsection.\thesubsection}{{\thesubsection} \parbox[t]{\dimexpr\textwidth-1pc}{#1}}}{1pc}{}

    \titleformat{name=\section,numberless}[display]{}{}{0pt}{\normalfont\Huge\bfseries #1}
\fi

\usepackage{cite} % Allows for ranges in citations
\usepackage{mciteplus}
\def\KstarzbFull {{\ensuremath{\Kbar{}^{*}(892)^{0}}}\xspace}
\def\DptoKKP {\ensuremath{\D^+\rightarrow K^-K^+\pi^+}\xspace}
\def\DptoKPP {\ensuremath{\D^+\rightarrow K^-\pi^+\pi^+}\xspace}

\def\DstoKKP {\ensuremath{\D^+_{\squark}\rightarrow K^-K^+\pi^+}\xspace}

\begin{document}

%%%%%%%%%%%%%%%%%%%%%%%%%
%%%%% Title     %%%%%%%%%
%%%%%%%%%%%%%%%%%%%%%%%%%
\renewcommand{\thefootnote}{\fnsymbol{footnote}}
\setcounter{footnote}{1}

% %%%%%%% CHOOSE TITLE PAGE--------
%\onecolumn
%\input{title-LHCb-INT}
%\input{title-LHCb-ANA}
%\input{title-LHCb-CONF}
%\input{title-LHCb-FIGURE}
% ===============================================================================
% Purpose: LHCb-PAPER journal paper title page template
% Author: 
% Created on: 2010-09-25
% ===============================================================================

%%%%%%%%%%%%%%%%%%%%%%%%%
%%%%%  TITLE PAGE  %%%%%%
%%%%%%%%%%%%%%%%%%%%%%%%%
\begin{titlepage}
\pagenumbering{roman}

% Header ---------------------------------------------------
\vspace*{-1.5cm}
\centerline{\large EUROPEAN ORGANIZATION FOR NUCLEAR RESEARCH (CERN)}
\vspace*{1.5cm}
\noindent
\begin{tabular*}{\linewidth}{lc@{\extracolsep{\fill}}r@{\extracolsep{0pt}}}
\ifthenelse{\boolean{pdflatex}}% Logo format choice
{\vspace*{-1.5cm}\mbox{\!\!\!\includegraphics[width=.14\textwidth]{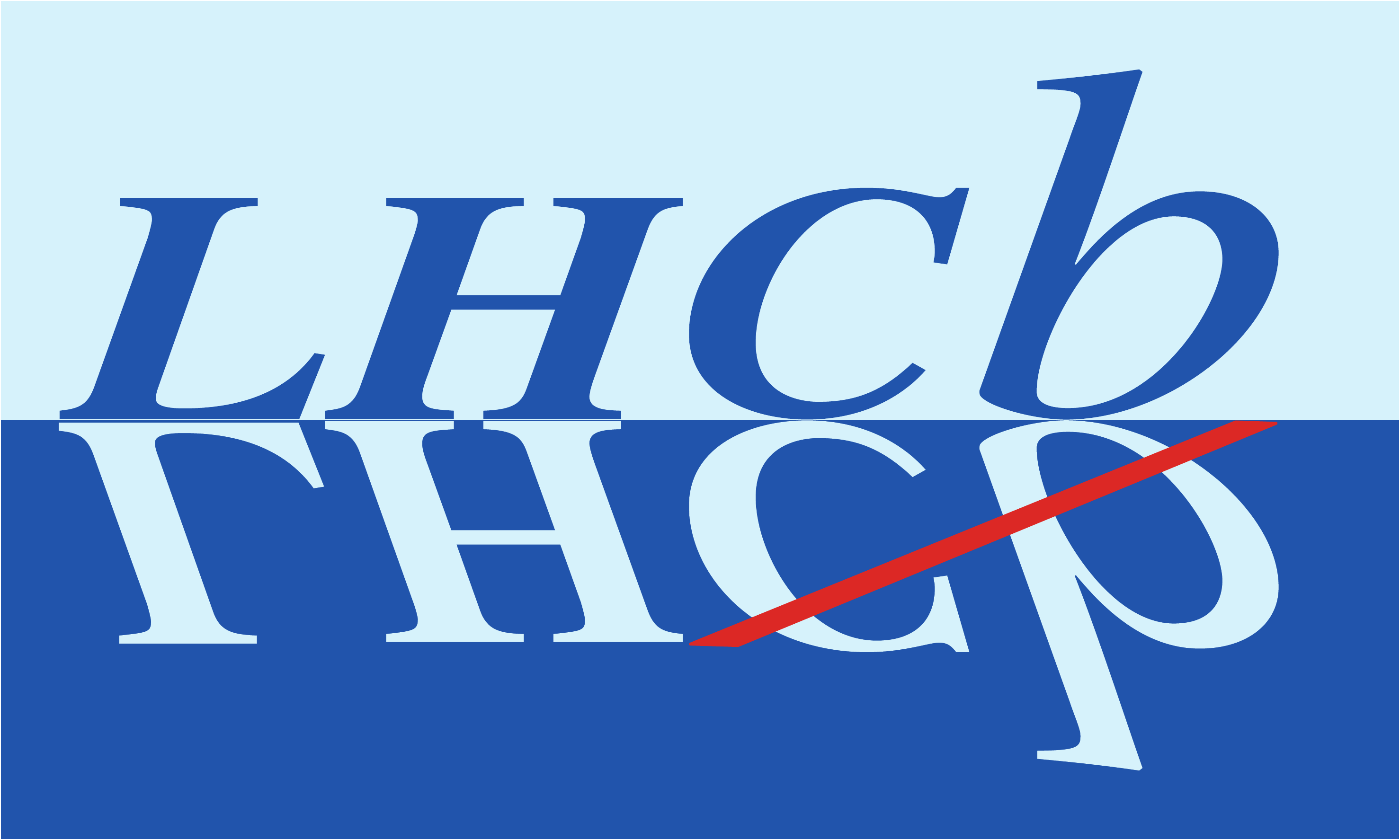}} & &}%
{\vspace*{-1.2cm}\mbox{\!\!\!\includegraphics[width=.12\textwidth]{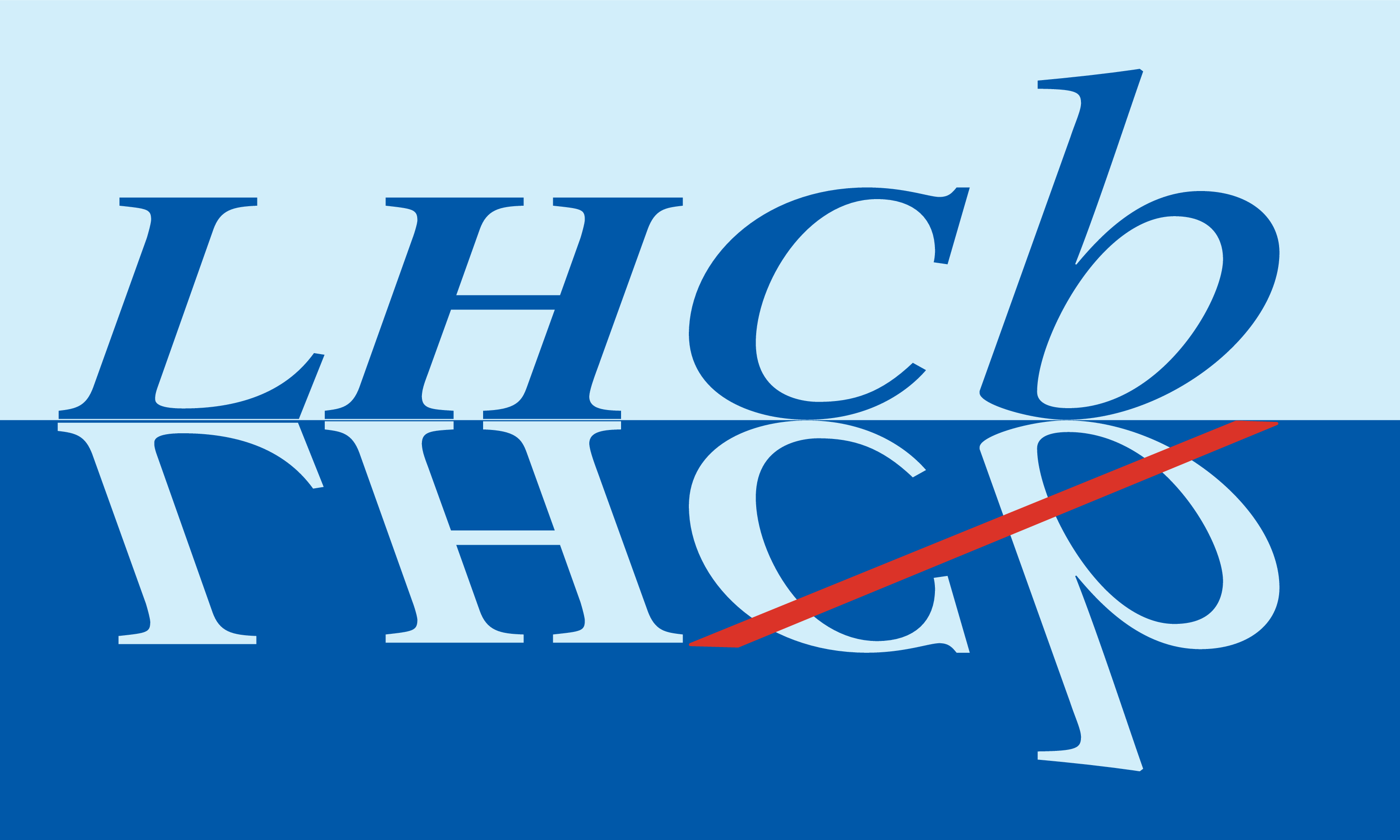}} & &}%
\\
 & & CERN-EP-2024-204 \\  % ID 
 & & LHCb-PAPER-2024-019 \\  % ID 
 & & February 4, 2025 \\ % Date - Can also hardwire e.g.: 23 March 2010
% & & v3.0 \\
% not in paper \hline
\end{tabular*}

\vspace*{3.5cm}

% Title --------------------------------------------------
{\normalfont\bfseries\boldmath\huge
\begin{center}
% DO NOT EDIT HERE. Instead edit macro in main.tex to keep metadata correct
  \papertitle 
\end{center}
}

\vspace*{1.0cm}

% Authors -------------------------------------------------
\begin{center}
%In the footnote, replace 'paper' by 'Letter' in case of submission to PRL or PLB 
% Edit macro in main.tex to keep metadata correct
\paperauthors\footnote{Authors are listed at the end of this Letter.}
\end{center}

\vspace{\fill}

% Abstract -----------------------------------------------
\begin{abstract}
  \noindent

	A search for violation of the charge-parity (${{\ensuremath{C\!P}}\xspace}$) symmetry in the \mbox{{\ensuremath{\D^+\rightarrow K^-K^+\pi^+}\xspace}} decay is presented, with proton-proton collision data corresponding to an integrated luminosity of 5.4{\ensuremath{\fb^{-1}}\xspace}, collected at a center-of-mass energy of $13\,\text{Te\kern -0.1em V}$ with the \mbox{LHCb}\xspace detector. A novel model-independent technique is used to compare the {{\ensuremath{\D^+}}\xspace} and {{\ensuremath{\D^-}}\xspace} phase-space distributions, with instrumental asymmetries subtracted using the \mbox{$\ensuremath{\D^+_{\ensuremath{s}\xspace}\rightarrow K^-K^+\pi^+}\xspace$} decay as a control channel. The $p$-value for the hypothesis of ${{\ensuremath{C\!P}}\xspace}$ conservation is $8.1\%$. The ${{\ensuremath{C\!P}}\xspace}$ asymmetry observables \mbox{$A_{{{\ensuremath{C\!P}}\xspace}|S}^{\phi\pi^+} = (0.95 \pm 0.43_{\rm stat} \pm 0.26_{\rm syst})\times 10^{-3}$} and \mbox{$A_{{{\ensuremath{C\!P}}\xspace}|S}^{\kern 0.18em\overline{\kern -0.18em K}^{*0}K^+} = (-0.26 \pm 0.56_{\rm stat} \pm 0.18_{\rm syst})\times 10^{-3}$} are also measured. These results show no evidence of {{\ensuremath{C\!P}}\xspace} violation and represent the most sensitive search performed through the phase space of a multibody decay.     
\end{abstract}

\vspace*{1.0cm}

\begin{center}
 Published in
  Phys.~Rev.~Lett. 133 (2024) 251801
\end{center}

\vspace{\fill}

{\footnotesize 
% Edit macro in main.tex to keep metadata correct
\centerline{\copyright~\papercopyright. \href{\paperlicenceurl}{\paperlicence}.}}
\vspace*{2mm}

\end{titlepage}

%%%%%%%%%%%%%%%%%%%%%%%%%%%%%%%%
%%%%%  EOD OF TITLE PAGE  %%%%%%
%%%%%%%%%%%%%%%%%%%%%%%%%%%%%%%%

%  empty page follows the title page ----
\newpage
\setcounter{page}{2}
\mbox{~}
%\newpage
%
%% Author List ----------------------------
%%  You need to get a new author list!
%\input{LHCb_authorlist.tex}
%
%The author list for journal publications is provided by the Membership Committee shortly after 'approval to go to paper' has been given.
%%It will be made available on the page
%%\verb!http://www.physik.uzh.ch/~strauman/forMemCo/LHCb-PAPER-XXXX-XXX/! .
%It will be sent to you by email shortly after a paper number has beens assigned.
%The author list should be included already at first circulation, 
%to allow new members of the collaboration to verify whether they have been included correctly.
%Occasionally a misspelled name is corrected or associated institutions become full members.
%In that case, a new author list will be sent to you.
%In case line numbering doesn't work well after including the authorlist, try moving the \verb!\bigskip! after the last author to a separate line.
%
%
%The authorship for Conference Reports should be ``The LHCb
%  collaboration'', with a footnote giving the name(s) of the contact
%  author(s), but without the full list of collaboration names.

%\twocolumn
% %%%%%%%%%%%%% ---------

\renewcommand{\thefootnote}{\arabic{footnote}}
\setcounter{footnote}{0}

%%%%%%%%%%%%%%%%%%%%%%%%%%%%%%%%
%%%%%  Table of Content   %%%%%%
%%%%%%%%%%%%%%%%%%%%%%%%%%%%%%%%
%%%% Uncomment if desired
%\tableofcontents

\cleardoublepage

%%%%%%%%%%%%%%%%%%%%%%%%%
%%%%% Main text %%%%%%%%%
%%%%%%%%%%%%%%%%%%%%%%%%%

\pagestyle{plain} % restore page numbers for the main text
\setcounter{page}{1}
\pagenumbering{arabic}

%% Uncomment during review phase. 
%% Comment before a final submission.
%\linenumbers

%% This is the main body
%% It is useful to have a single file so comments are not missed in overleaf.
The breaking of the combined charge-parity (\CP) symmetry in particle interactions is a critical element in explaining the imbalance between matter and antimatter in the Universe~\cite{Sakharov:1967}. Existing observations of \CP violation in \kaon and \B meson decays are consistent with predictions from the Standard Model (SM), where the violation arises from a single irreducible phase in the Cabibbo--Kobayashi--Maskawa (CKM) quark mixing matrix~\cite{Cabibbo:1963yz,Kobayashi:1973fv}. However, the effect of this phase alone is insufficient to explain the observed matter-antimatter asymmetry~\cite{Gavela:1994dt,Gavela:1993ts, Cohen1993ProgressIE,annurev.nucl.49.1.35}, underscoring the necessity for new \CP-violating dynamics beyond the SM.

 In the charm-quark sector, \CP violation has so far only been observed in the difference between the time-integrated \CP asymmetries of \mbox{\decay{\Dz}{\pim\pip}} and \decay{\Dz}{\Km\Kp} decays~\cite{LHCb-PAPER-2019-006}.  These are Cabibbo-suppressed decays, with transitions of the type $c\to u q \bar q$ ($q=d,s$)  for which the SM predicts very small  \CP asymmetries, of the order of $10^{-3}$ or less. The recent measurement of the asymmetry in the  $\decay{\Dz}{\Km\Kp}$ decay~\cite{LHCb-PAPER-2022-024,LHCb-CONF-2024-004,*LHCb-PAPER-2021-033} indicates that the observed signal is mostly due to \CP violation in the $\decay{\Dz}{\pim\pip}$ decay amplitude. The interpretation of these results is still being debated, with explanations falling within the SM~\cite{Grossman:2019xcj,Cheng:2019ggx,Schacht:2021jaz,Bediaga:2022sxw} and some suggesting physics beyond it~\cite{Grossman:2006jg,Khodjamirian:2017zdu,Pich:2023kim,Lenz:2023rlq}. Measurements of \CP asymmetries in a wide range of decays are fundamental to fully exploit the potential of charmed hadrons for revealing signs of physics beyond the SM.

Three-body decays have unique features for \CP-asymmetry studies.  Depending on the underlying mechanism of \CP violation, localized asymmetries across the decay phase-space can be considerably larger than the integrated value, as observed already in \Bp decays~\cite{LHCb-PAPER-2019-018,LHCb-PAPER-2021-049}. The \DptoKKP decay\footnote{The inclusion of charge-conjugate processes is implied throughout except in the discussion of asymmetries.}  is the three-body Cabibbo-suppressed \Dp decay with the largest branching fraction~\cite{PDG2022}. Studies of \CP violation across its Dalitz plot, a two-dimensional diagram representing the phase space~\cite{Dalitz:1953cp,Fabri:1954zz}, have previously been undertaken by the \babar~\cite{BaBar:2005enq}, \cleo~\cite{CLEO-PhysRevD.78.072003} and  \lhcb~\cite{LHCb-PAPER-2011-017} collaborations. Measurements of the \CP asymmetry in the phase-space region dominated by $\decay{\Dp}{\phiz\pip}$, with $\decay{\phiz}{\Km\Kp}$, have also been reported by \lhcb~\cite{LHCB-PAPER-2012-052,LHCb-PAPER-2019-002}. All results are consistent with \CP symmetry.

In this Letter, a search for localized \CP violation in the phase space of \DptoKKP decays is presented, using data collected by the \lhcb experiment from 2016 to 2018, corresponding to an integrated luminosity of 5.4\invfb. 
Currently, this represents the largest sample available of $\cquark$-hadron decays with potential for observation of \CP violation.  A model-independent technique is developed to address the challenge of isolating the \CP asymmetry from instrumental effects, by exploiting the control channel \DstoKKP, for which \CP violation is not expected~\cite{Grossman:2006jg, Bediaga:2020}. 

The LHCb detector is a single-arm forward spectrometer covering the pseudorapidity range $2 < \eta < 5$, 
described in detail in Refs.~\cite{LHCb-DP-2008-001,LHCb-DP-2014-002}. It is designed for the study of particles containing \bquark\ or \cquark\ quarks. 
The magnetic field polarity is reversed periodically during data taking to mitigate the differences of reconstruction efficiencies of particles with opposite charges. Data sets corresponding to about one half of the total integrated luminosity are recorded with each magnetic field configuration. 

The online event selection is performed by a trigger consisting of a hardware stage followed by a two-level software stage. In between the two software stages, an alignment and calibration of the detector is performed in near real-time and their results are used in the trigger~\cite{LHCb-PROC-2015-011}.
In the first software stage, events used in this analysis are selected if at least one track has large transverse momentum and is incompatible with originating
from any primary vertex (PV), or if any two-track combination forming a vertex significantly displaced from the PV is found in the event by a multivariate algorithm\cite{LHCb-DP-2012-004,LHCb-DP-2016-001}. In the second stage, two oppositely charged particles identified as kaons are combined with a particle identified as a pion to form a good-quality decay vertex detached from any PV. The PV with the smallest value of $\chisq_{\rm IP}$ is associated with the decay candidate, where $\chisq_{\rm IP}$ is defined as the difference in the vertex-fit \chisq of the PV reconstructed with and without the particle under consideration, in this case, the \DporDsp candidate. Further requirements are applied on the \DporDsp decay time, on the angle between the reconstructed \DporDsp momentum vector and the vector connecting the PV to the decay vertex, on the  $\chisq$ of the \DporDsp decay-vertex fit, and on the momentum, the transverse momentum and the $\chisq_{\rm IP}$ of the \DporDsp candidate and of its decay products. The invariant masses of the \Dp and \Dsp candidates are required to be within the intervals \mbox{[1805,1935]}\mev and \mbox{[1905,2035]}\mev, which correspond to approximately ten times their mass resolution.\footnote{Natural units with $c = 1$ are used throughout.}

The candidates selected from the second software-trigger stage are directly used for the offline analysis~\cite{LHCb-DP-2016-001} and must satisfy additional selection criteria, applied uniformly for the \Dp and \Dsp samples whenever possible, to minimize potential biases in the \CP violation study.    A stringent particle identification (PID) requirement is applied on both kaons, which reduces to a negligible level the contamination from \mbox{\DptoKPP} decays in the \DstoKKP sample, while significantly reducing background due to random combinations of tracks.
Contamination from semileptonic decays where the muon is misidentified as a pion is suppressed by a muon PID veto to the pion candidate. Contamination from $\decay{\Lc}{\Km\proton\pip}$ decays and random combinations of $\decay{\Dz}{\Km\pip}$ decays with a pion are removed using invariant-mass vetoes. 
The requirement on the maximum $\chisq_{\rm IP}$ of the $\DporDsp$ meson is further tightened to reduce the contamination from secondary \bquark-hadron decays. 
Fiducial requirements are applied to remove small regions of the kinematic space where large charge asymmetries are observed, typically caused by low-momentum tracks being swept out of the detector acceptance by the magnetic field. Finally, candidates are randomly rejected to equalize, in each data-taking year, the signal yields of the samples collected with each magnet polarity.  

After applying all selection criteria, the yields for \Dp and \Dsp decays are approximately 135 and 181 million, respectively, with a purity of 95\% within an interval of $\pm 20 \mev$ around their nominal masses. The $\Km\Kp\pip$ invariant-mass distributions are shown in Fig.~\ref{fig:mkkpi} of the Appendix. 
 
 The strategy to identify a signal of \CP violation in the \DptoKKP decay is a model-independent search based upon a \chisq-test for the hypothesis of the charge asymmetries over the \Dp Dalitz plot being compatible with those of the \Dsp control mode. It assumes that the nuisance asymmetries (\ie, differences in charged-particle detection, reconstruction or selection efficiencies, left-right detector or any other instrumental asymmetries) depend only on the kinematics of the final-state particles and that local effects arising from correlations between the \DporDsp kinematic-dependent production asymmetries and the Dalitz plot variables are negligible, making the difference between the production asymmetries a global effect that can be subtracted. 

The Dalitz plots are divided into bins and the raw charge asymmetry for the decay $X$ in each bin $i$ is defined as
 \begin{equation}
A_{\rm raw}^{\, i, X} = \frac{N_+^{\, i, X} - N_-^{\, i, X}}{N_+^{\, i, X} + N_-^{\, i, X}}, 
 \end{equation}
where $X = S, C$ refers to the signal ($S$) or control ($C$) modes, $+$ and $-$ refer to the charge of the $\DpmorDspm$ candidates, and $N^{\, i, X}_{+(-)}$ is the number of decays $X_{+(-)}$ in bin $i$, as determined from mass fits described later. The \CP observable is defined as  
 \begin{equation}
 \label{eq:deltaAi}
  \Delta A^{\, i}_{\CP}  =  A_{\rm raw}^{\, i, S} - A_{\rm raw}^{\, i, C} - \Delta A_{\rm raw}^{\text{global}}. 
 \end{equation}
 The term  $\Delta A_{\rm raw}^{\text{global}}$ is the  global difference in asymmetries averaged  over all bins in the Dalitz plot, 
 \begin{equation}
 \Delta A_{\rm raw}^{\text{global}} = \frac{
 \displaystyle\sum_i^{N_{\text{bins}}} \frac{ A_{\rm raw}^{\, i, S}-A_{\rm raw}^{\, i, C}}{\sigma^2_{ A_{\rm raw}^{\, i, S}}+\sigma^2_{ A_{\rm raw}^{\, i, C}}}}{\displaystyle\sum_i^{N_{\text{bins}}} \frac{1}{\sigma^2_{ A_{\rm raw}^{\, i, S}}+\sigma^2_{A_{\rm raw}^{\, i, C}}}},
  \end{equation}
  where $\sigma_{ A_{\rm raw}^{\, i, X}}$ are the $ A_{\rm raw}^{\, i, X}$  uncertainties. Any difference between the global asymmetries in the two decays is cancelled by this term 
  which, in the absence of \CP violation, corresponds to the difference between the \Dp and \Dsp production asymmetries. The $ A_{\rm raw}^{\, i, C}$ term cancels out instrumental asymmetries from the raw signal asymmetry in bin $i$.

To build the \chisq test-statistic,  the significance of $\Delta A^i_{\CP}$, 
\begin{equation}
\mathcal{S}_{\Delta_{\CP}}^{\, i}=\frac{\Delta A^i_\CP}{\sigma_{\Delta A^i_\CP}},    
\label{eq:SDeltaCP}
\end{equation} 
is used, where $\sigma_{\Delta A^i_\CP}$ is 
 the $\Delta A^i_{\CP}$ statistical uncertainty 
and accounts for the correlations between the individual per-bin asymmetries and the global difference of asymmetries. The expressions for  $\sigma_{\Delta A^i_\CP}$ and $\sigma_{\Delta A^{_{\rm raw}^{\text{global}}}}$ are presented in Eq.~\ref{eq:sigma_DeltaAi} and Eq.~\ref{eq:sigma_DeltaA_global} in the Appendix. Under the hypothesis of \CP symmetry, $\mathcal{S}_{\Delta_{\CP}}^{\, i}$ follows a standard normal distribution, as validated using pseudoexperiments.

The test-statistic is then defined as
\begin{equation}
\chisq(\mathcal{S}_{\Delta_{\CP}}) = \displaystyle\sum_i^{N_{\text{bins}}}  (\mathcal{S}_{\Delta_{\CP}}^{\, i})^2,
\end{equation}
from which a $p$-value for the hypothesis of no localized \CP violation can be extracted.

\begin{figure*}[tp]
\begin{center}
       \includegraphics[width=\textwidth]{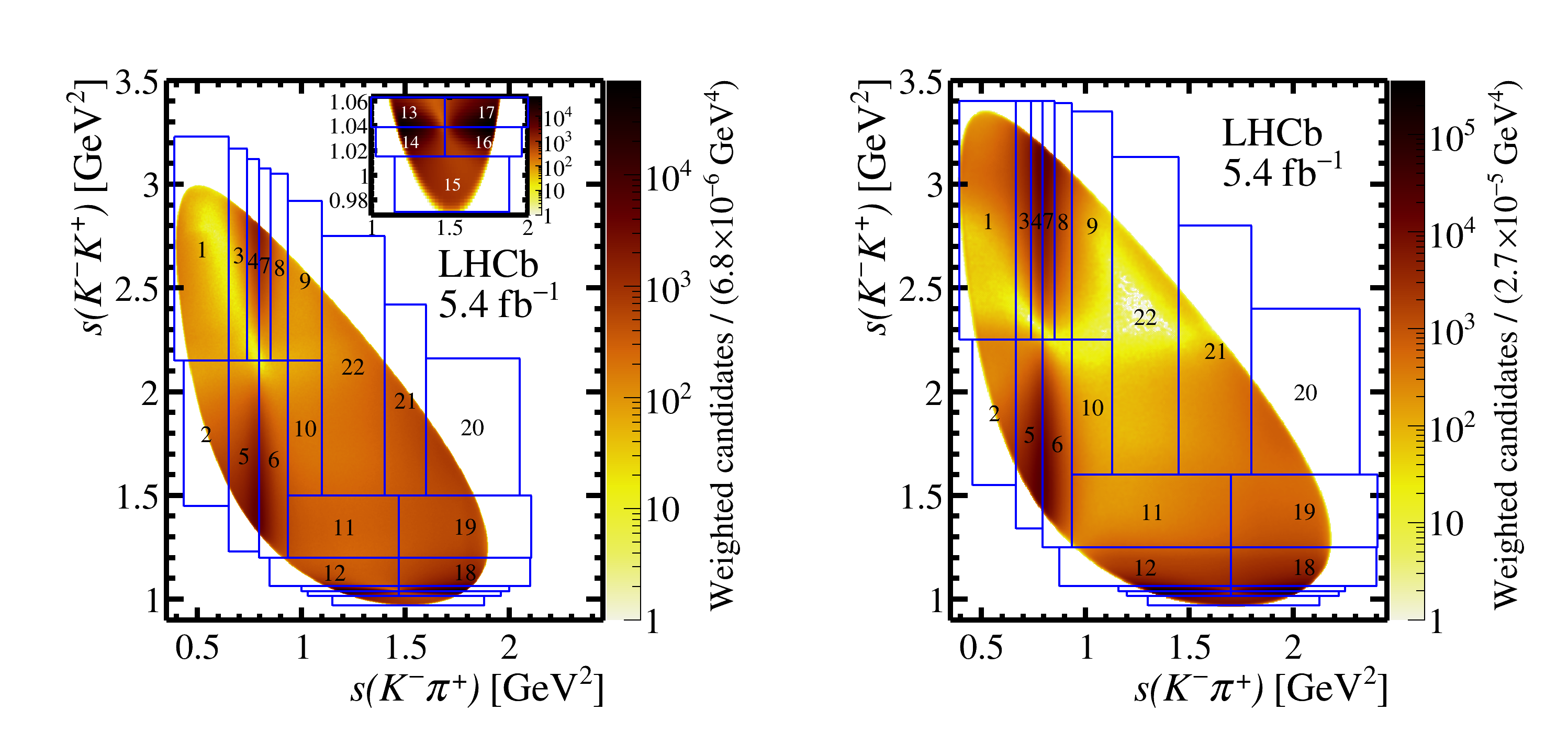}      
 \caption{Dalitz plots for (left) $\DptoKKP$ and (right) $\DstoKKP$ decays in data with the binning scheme overlaid. The enlarged inset shows the bins around the $\phiz\pip$ region, where the same numbering scheme is followed for both channels.} \label{fig:DP} 
 \vspace*{-0.45cm}
  \end{center}
 \end{figure*}

The Dalitz plots for the signal and control samples are depicted in Fig.~\ref{fig:DP}, where the binning scheme is overlaid. In this plot, $s(\Km\pip)$ and $s(\Km\Kp)$ represent  the squared invariant masses of the $\Km\pip$ and $\Km\Kp$ systems, respectively. For the purpose of this plot, background subtraction is performed using the \sPlot technique~\cite{Pivk:2004ty},  based on \Dp and \Dsp invariant-mass fits.

The binning scheme is designed to exploit the resonant structures in the \mbox{\DptoKKP} decay, which are similar to those in the \DstoKKP decay. Key features in these decays are the  $\KstarzbFull$ and $\phiz(1020)$ vector resonances,  prominently visible around  $s(\Km\pip)=m^2_{\Kstarzb}$ and $s(\Km\Kp)=m^2_{\phi}$. Vector resonance distributions in the Dalitz plot present a node, shown more explicitly in the inset of Fig.~\ref{fig:DP} for the $\phiz$ case. The rectangular areas around the two resonances are divided into four bins with the internal boundaries defined by the distribution node in one direction and the known mass~\cite{PDG2022} in the other. For the \Kstarz resonance, the region above the node is further subdivided to account for instrumental asymmetry variations. Other Dalitz plot areas are segmented to approximately equalize the number of candidates in each bin, with the \Dsp Dalitz plot following a similar segmentation with expanded outer limits. This design has a natural bin-to-bin correspondence between the Dalitz plots of \Dp and \Dsp mesons with a good agreement between final-state kinematic quantities in each bin. A bin-by-bin weighting procedure is applied to the \Dsp candidates to correct for residual differences.

Sensitivity studies, where \DptoKKP samples are generated according to the isobar model from Ref.~\cite{CLEO-PhysRevD.78.072003}, support this binning choice. In these studies, \CP violation is introduced as small differences in the relative phase or magnitude of one or more resonances' amplitudes between the \Dp and \Dm samples. The asymmetry caused by different phases in a given resonant contribution to the \Dp and \Dm amplitudes changes sign when crossing vertically or horizontally between quadrants around this resonance. This behavior motivates the measurement of the \CP asymmetry observable $A_{\CP|S}$\cite{LHCB-PAPER-2012-052}:
 \begin{equation}
    A_{\CP|S} = \frac{1}{2}\left[\left(\Delta A_{\rm raw}^\text{top-left}+\Delta A_{\rm raw}^\text{bottom-right}\right)-\left(\Delta A_{\rm raw}^\text{top-right}+\Delta A_{\rm raw}^\text{bottom-left}\right)\right], 
 \end{equation}
where $\Delta A_{\rm raw}=  A_{\rm raw}^{S} - A_{\rm raw}^{C}$ and the Dalitz plot bins top-left, top-right, bottom-left, and bottom-right are the ones numbered in Fig.~\ref{fig:DP} as bins 13, 17, 14 and 16 (3+4, 7+8, 5 and 6),\footnote{The plus sign indicates that candidates from two bins are combined before computing $A_{\CP|S}$.} respectively, for the $\phiz\pip$ ($\Kstarzb\Kp$) resonant amplitude. In this case, the global asymmetry components cancel out, making $ A_{\CP|S}$ an observable sensitive purely to \CP violation in the decay. The local $\Delta A^i_{\CP}$ are also provided as additional measurements, which can be interpreted as \CP-violating observables in light of theoretical predictions, but only as relative quantities between bins. The net average \CP asymmetry is removed in the subtraction, and therefore, the $\Delta A^i_{\CP}$ in individual bins cannot be combined to obtain the integrated \CP asymmetry over the phase space. All the results from the present study are insensitive to \CP violation manifested solely as a global asymmetry. 
 
To verify the method and mitigate experimenter's bias, multiple tests were conducted before examining the $A_{\rm raw}^{\,i,S}$ values. Both signal and control samples were segmented by \DporDsp momentum, pseudorapidity, and transverse momentum, confirming that $\Delta A^i_{\CP}$ values were statistically compatible among these segments.  Simulations \cite{Cowan:2016tnm} of the \DptoKKP and \DstoKKP decays, using a realistic resonant model with the same yields as in data, are used to study the impact of known effects, such as detection and tracking efficiency asymmetries~\cite{Davis:2310213} and PID efficiency asymmetries extracted from calibration samples~\cite{LHCb-DP-2018-001}, obtained as a function of the kaon and pion kinematics. The patterns of asymmetries in simulated signal and control samples match those observed in the control data.  Additionally, various production-asymmetry models are tested, based on the results obtained with data taken in 2011 and 2012 \cite{LHCb-PAPER-2012-026, LHCb-PAPER-2012-009, LHCb-PAPER-2018-010}. None of the models significantly affects the local raw asymmetry patterns across the Dalitz plot. The method's validity is further confirmed through 10,000 pseudoexperiments designed to simulate instrumental asymmetries, with the resulting $\sum_i^{N_{\text{bins}}} (\mathcal{S}_{\Delta{\CP}}^i)^2$ values fitting a $\chisq$ distribution with $N_{\text{bins}}$ degrees of freedom.  

The signal and background raw charge asymmetries and yields are directly determined for each Dalitz plot bin from simultaneous \chisq fits to the binned $K^\mp K^\pm \pi^\pm$ invariant-mass distributions of \DporDsp and \DmorDsm candidates, independently for each data taking year and magnet polarity. The raw asymmetries are then combined for the final results. The signal and background shapes, as well as the procedure used to verify that the extracted asymmetries are unbiased, are described in the Appendix.

The results for  $\Delta A^i_{\CP}$ are shown in detail in Table~\ref{tab:syst_DeltaACP}.
Systematic uncertainties arise from various sources and are evaluated by comparing results obtained with alternative procedures to the nominal ones. The impact of remaining kinematic mismatches between the control and signal samples is assessed by obtaining the \Dsp asymmetries without equalization weights. Different fit models for the signal and background mass shapes are tested. Biases due to differences in the instrumental asymmetries related to the different \Dp and \Dsp lifetimes, flight distance distributions and contamination from $\bquark$-hadron decays, which cannot be disentangled, are determined by splitting the samples into two ranges of $\D$-meson flight distance significance and $\chisq_{\rm IP}$ and combining the resulting $\Delta A^i_{\CP}$ values. For most bins, this is the dominant systematic uncertainty (labelled as ``sec.dec." in Table~\ref{tab:syst_DeltaACP}). The influence of different requirements in the trigger for the signal and control channels is investigated by aligning the selection criteria. The precision in each bin is limited by the statistical uncertainty. A graphical representation of the results is shown in Fig.~\ref{fig:DeltaACP} of the Appendix.

\begin{table}[ht]
\begin{center}
\caption{Experimental results for $\Delta A_{\CP}^i = \Delta A_{\text{raw}}^i - \Delta A_{\text{raw}}^{\text{global}}$ and $A_{\CP|S}$, with systematic and statistical uncertainties shown in the last two columns. The individual contributions to the systematic uncertainties are also shown. All values are expressed in $10^{-3}$ units.}
\label{tab:syst_DeltaACP}
\begin{tabular}{S[table-format=2.2] S[table-format=2.2] S[table-format=1.2] S[table-format=1.2] S[table-format=1.2] S[table-format=1.2] S[table-format=1.2] S[table-format=1.2]}
\toprule
{Bin} & {$\Delta A_{\CP}$} 
& {$\sigma_{\rm{syst}}^{\rm{weighting}}$} & {$\sigma_{\rm{syst}}^{\rm{sec. dec.}}$} & {$\sigma_{\rm{syst}}^{\rm{fit}}$} & {$\sigma_{\rm{syst}}^{\rm{trigger}}$} & {$\sigma^{\rm{tot}}_{\rm{syst}}$} & {$\sigma_{\rm{stat}}$} \\
\midrule
1   & 1.97 & 0.67 & 1.35 & 0.88 & 0.10 & 1.75 & 1.39 \\
2   & 0.78 & 0.32 & 0.19 & 0.50 & 0.26 & 0.68 & 1.33 \\
3   & -0.01 & 0.01 & 0.49 & 0.15 & 0.67 & 0.85 & 1.43 \\
4   & 0.06 & 0.20 & 0.11 & 0.24 & 0.21 & 0.39 & 0.75 \\
5   & 0.24 & 0.06 & 0.07 & 0.04 & 0.05 & 0.11 & 0.36 \\
6   & 0.71 & 0.05 & 0.21 & 0.08 & 0.17 & 0.29 & 0.38 \\
7   & 0.99 & 0.06 & 0.36 & 0.19 & 0.04 & 0.41 & 0.73 \\
8   & 0.69 & 0.16 & 0.57 & 0.13 & 0.13 & 0.63 & 1.28 \\
9   & -3.14 & 0.99 & 0.60 & 1.59 & 0.01 & 1.97 & 2.49 \\
10  & 0.83 & 1.28 & 1.97 & 2.63 & 0.87 & 3.63 & 1.97 \\
11  & -0.02 & 0.20 & 0.39 & 0.10 & 0.24 & 0.51 & 0.78 \\
12  & 1.08 & 0.11 & 0.22 & 0.11 & 0.03 & 0.27 & 0.63 \\
13  & 0.25 & 0.04 & 0.13 & 0.06 & 0.06 & 0.17 & 0.42 \\
14  & -0.99 & 0.08 & 0.05 & 0.07 & 0.08 & 0.14 & 0.42 \\
15  & -1.99 & 0.01 & 0.56 & 0.08 & 0.18 & 0.60 & 0.90 \\
16  & 0.12 & 0.05 & 0.26 & 0.07 & 0.15 & 0.31 & 0.43 \\
17  & -0.55 & 0.08 & 0.08 & 0.10 & 0.03 & 0.15 & 0.35 \\
18  & -0.24 & 0.19 & 0.39 & 0.03 & 0.14 & 0.46 & 0.47 \\
19  & -0.53 & 0.07 & 0.50 & 0.20 & 0.34 & 0.64 & 0.75 \\
20  & -2.14 & 0.46 & 0.29 & 0.29 & 0.20 & 0.65 & 1.14 \\
21  & 0.45 & 0.48 & 0.53 & 0.23 & 0.42 & 0.86 & 1.07 \\
22  & 2.14 & 1.28 & 1.09 & 0.81 & 0.85 & 2.05 & 2.23 \\
\bottomrule
{Resonant mode} & {$A_{\CP|S}$} & {$\sigma_{\rm{syst}}^{\rm{weighting}}$} & {$\sigma_{\rm{syst}}^{\rm{sec.dec.}}$} & {$\sigma_{\rm{syst}}^{\rm{fit}}$} & {$\sigma_{\rm{syst}}^{\rm{trigger}}$} & {$\sigma^{\rm{tot}}_{\rm{syst}}$} & {$\sigma_{\rm{stat}}$} \\
\midrule
$\phi\pip$ & 0.95  & 0.12 & 0.21 & 0.07 & 0.07 & 0.26 & 0.43 \\
$\Kstarzb\Kp$ & -0.26 & 0.01 & 0.16 & 0.06 & 0.06 & 0.18 & 0.56 \\
\bottomrule
\end{tabular}
\end{center}
\end{table}

 The \CP observables ${A}_{\CP|S}$ for the $\phi\pip$ and $\Kstarzb\Kp$ systems are  given by
\begin{align*}
    A_{\CP|S}^{\phi\pip} &= (0.95 \pm 0.43 \pm 0.26)\times 10^{-3}, \\
    A_{\CP|S}^{\Kstarzb\Kp} &= (-0.26 \pm 0.56 \pm 0.18)\times 10^{-3},
\end{align*}
where the first uncertainty is statistical and the second is systematic,  evaluated using the same procedure described for the $\Delta A^i_{\CP}$ results. The uncertainties are also shown in detail in Table~\ref{tab:syst_DeltaACP}.

\begin{figure}[h]
\begin{center}      \includegraphics[width=0.65\textwidth]{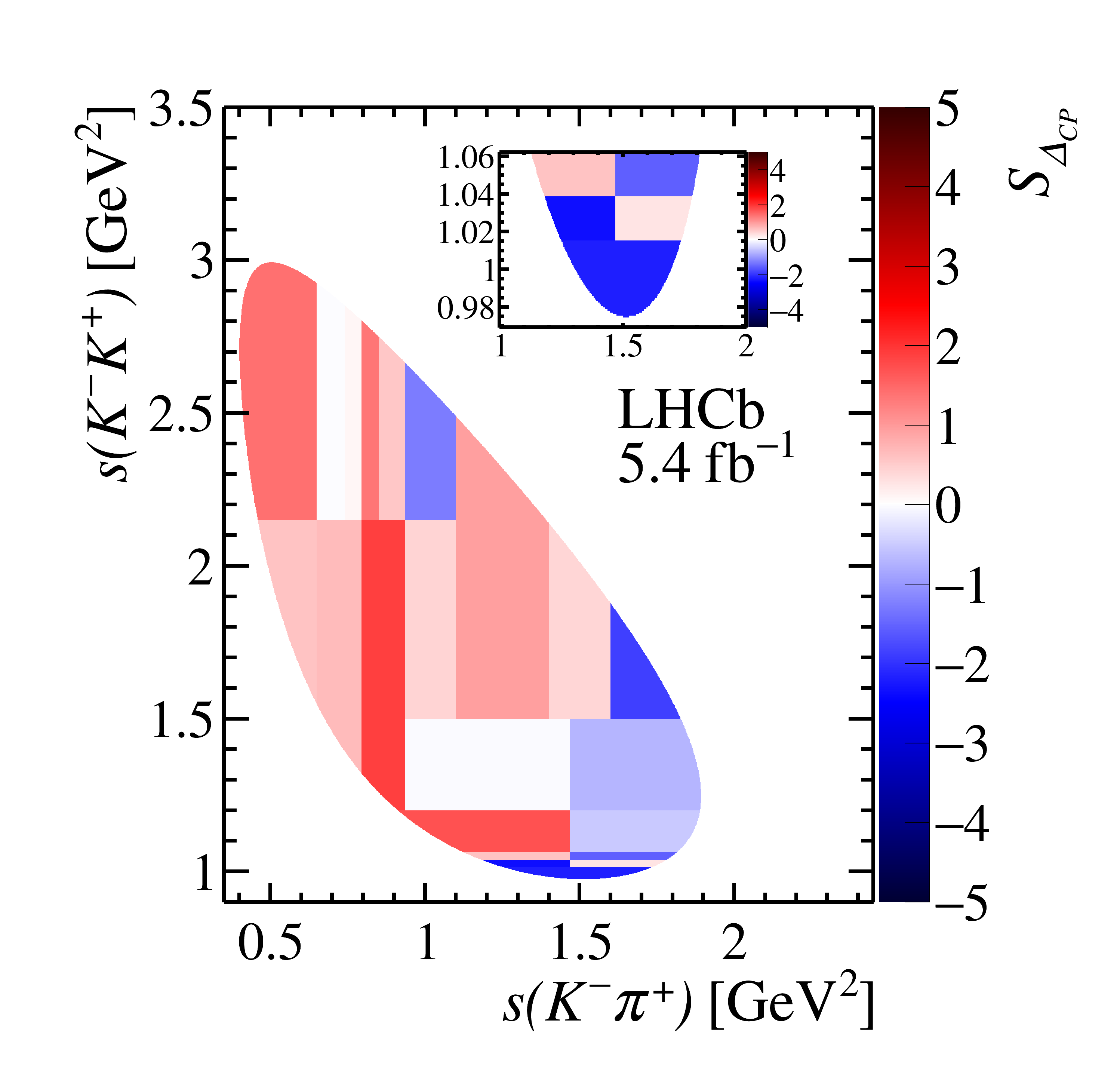}
\end{center}
    \caption{The significance $\mathcal{S}_{\Delta_{\CP}}$ across the Dalitz plot, which accounts only for the statistical uncertainty. The inset shows a zoom of the Dalitz plot around the $\phiz\pip$ region.}
    \label{fig:sdcp}
\end{figure}

Finally, the statistical significances $\mathcal{S}_{\Delta_{\CP}}^{\, i}$ are shown across the Dalitz plot in Fig.~\ref{fig:sdcp}.  The $\chisq$ of the search test is 31.8 for 22 degrees of freedom, resulting in a $p$-value of 8.1$\%$, consistent with the hypothesis of no localized \CP violation in the phase space of the \DptoKKP decay. Potential systematic biases in the search test results are examined by evaluating $\Delta A^i_{\CP}$ in the same alternative scenarios used to evaluate systematic uncertainties for $\Delta A_{\CP}^{i}$. The resulting $p$-values range from 2.3\% to 14.1\%, confirming absence of localized \CP violation over the Dalitz plot within the current statistical reach.

In summary, this Letter reports a search for \CP violation across the phase space of \DptoKKP decays utilizing a novel binned, model-independent method that neutralizes instrumental effects via a control channel with an identical final state. The $p$-values,  obtained from all considered scenarios,  are consistent with  \CP symmetry. The observables $\Delta A^i_{\CP}$ are measured with  $10^{-3}$ precision within the bins of the Dalitz plot. This constitutes the most precise search for localized  \CP violation ever performed. Furthermore, this study presents the most precise measurement of $A_{\CP|S}^{\phi\pip}$ and the first experimental result for $A_{\CP|S}^{\Kstarzb\Kp}$, both compatible with zero with a precision smaller than $10^{-3}$.

% Do not include this in any draft (just for information in the template)
%\input{acknowledgements_intro}
% Comment this in for paper drafts; do not include this in analysis note, conference and figure reports
\newpage
\section*{Acknowledgements}
%
% These Acknowledgements valid from 3-May-2019
%
\noindent We express our gratitude to our colleagues in the CERN
accelerator departments for the excellent performance of the LHC. We
thank the technical and administrative staff at the LHCb
institutes.
We acknowledge support from CERN and from the national agencies:
CAPES, CNPq, FAPERJ and FINEP (Brazil); 
MOST and NSFC (China); 
CNRS/IN2P3 (France); 
BMBF, DFG and MPG (Germany); 
INFN (Italy); 
NWO (Netherlands); 
MNiSW and NCN (Poland); 
MCID/IFA (Romania); 
%MSHE (Russia); 
MICIU and AEI (Spain);
SNSF and SER (Switzerland); 
NASU (Ukraine); 
STFC (United Kingdom); 
DOE NP and NSF (USA).
We acknowledge the computing resources that are provided by CERN, IN2P3
(France), KIT and DESY (Germany), INFN (Italy), SURF (Netherlands),
PIC (Spain), GridPP (United Kingdom), 
%RRCKI and Yandex LLC (Russia), 
CSCS (Switzerland), IFIN-HH (Romania), CBPF (Brazil),
and Polish WLCG (Poland).
We are indebted to the communities behind the multiple open-source
software packages on which we depend.
Individual groups or members have received support from
ARC and ARDC (Australia);
Key Research Program of Frontier Sciences of CAS, CAS PIFI, CAS CCEPP, 
Fundamental Research Funds for the Central Universities, 
and Sci. \& Tech. Program of Guangzhou (China);
Minciencias (Colombia);
EPLANET, Marie Sk\l{}odowska-Curie Actions, ERC and NextGenerationEU (European Union);
A*MIDEX, ANR, IPhU and Labex P2IO, and R\'{e}gion Auvergne-Rh\^{o}ne-Alpes (France);
%RFBR, RSF and Yandex LLC (Russia);
AvH Foundation (Germany);
ICSC (Italy); 
%GVA, XuntaGal, GENCAT, Inditex, InTalent and Prog.~Atracci\'on Talento, CM (Spain);
Severo Ochoa and Mar\'ia de Maeztu Units of Excellence, GVA, XuntaGal, GENCAT, InTalent-Inditex and Prog. ~Atracci\'on Talento CM (Spain);
SRC (Sweden);
the Leverhulme Trust, the Royal Society
 and UKRI (United Kingdom).

% This should be taken out in the final paper
\appendix

\section{Appendix}
\label{sec:Supplementary}

\subsection{Invariant-mass distributions and fits}

The probability density function (PDF) used to describe the $K^\mp K^\pm \pi^\pm$  invariant-mass distribution of signal candidates in each bin of the Dalitz plot is given by a sum of a Gaussian function and a double-sided Crystal Ball (DSCB) function~\cite{Skwarnicki:1986xj}, with widths and tails that can differ between the left and right sides of the PDF around the most probable value. All the parameters describing the signal shape, apart from the Gaussian width and mean, are fixed to the values obtained from simulation, weighted by PID efficiencies obtained from calibration samples~\cite{LHCb-DP-2018-001}. The fixed parameters are the offset and relative fraction between the DSCB and the Gaussian functions, and the tail parameters of the DSCB. The background is parameterized by a third-order Bernstein polynomial for \DptoKKP and as a second-order for polynomial for \DstoKKP decays. The Gaussian width and the background parameters are shared between the \DporDsp and \DmorDsm subsamples that are fitted simultaneously.

The fits are validated using pseudoexperiments, where $K^\mp K^\pm \pi^\pm$ invariant-mass distributions are generated according to the baseline fit results and then fitted with the same prescription. The pulls for the fitted parameters follow standard normal distributions across all Dalitz plot bins for both decay modes. This analysis confirms that the signal and background asymmetries are unbiased.

\begin{figure*}[btp]
\begin{center}
 \begin{tabular}{cc}
        \includegraphics[width=0.47\textwidth]{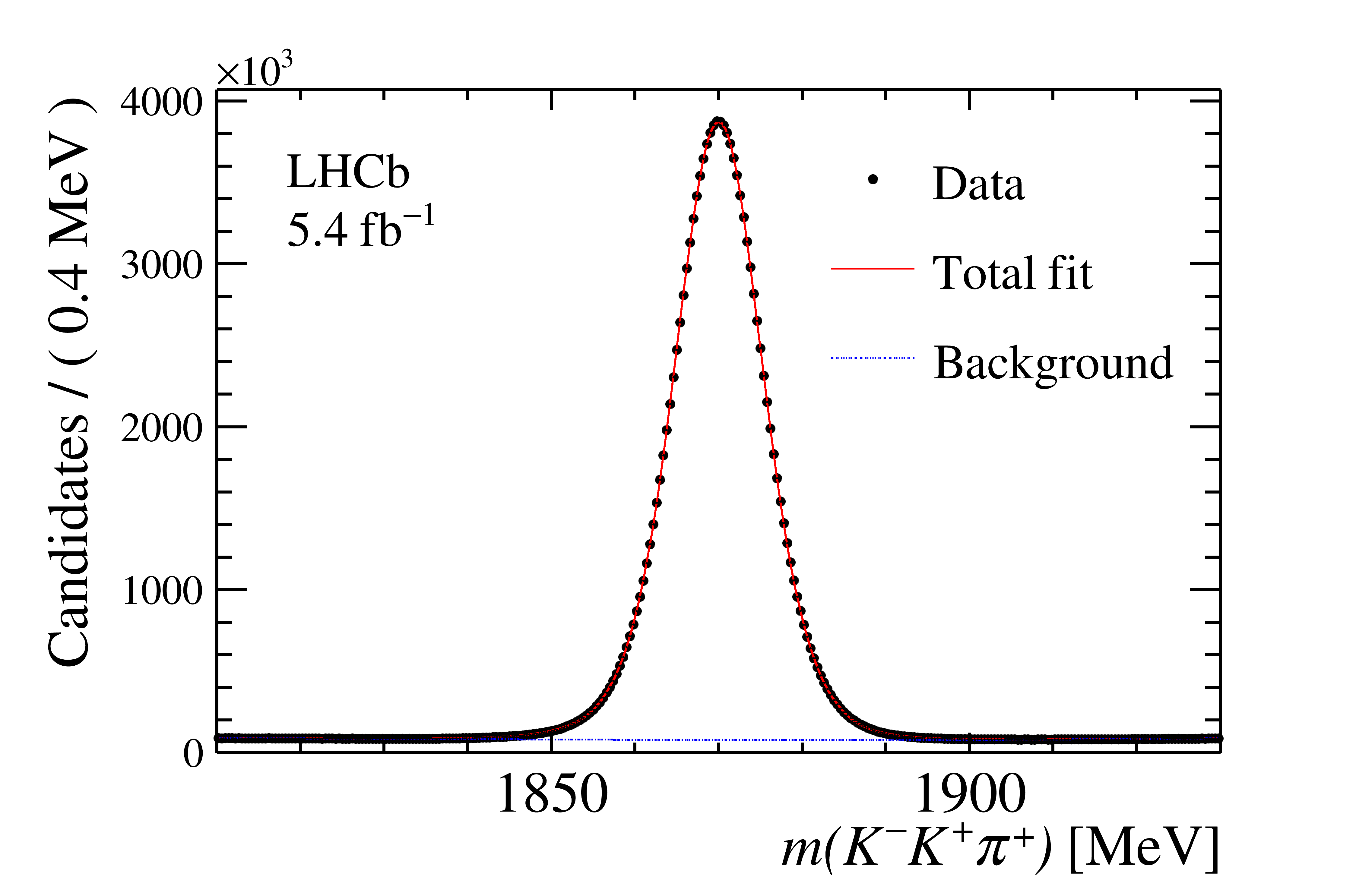} & 
         \includegraphics[width=0.47\textwidth]{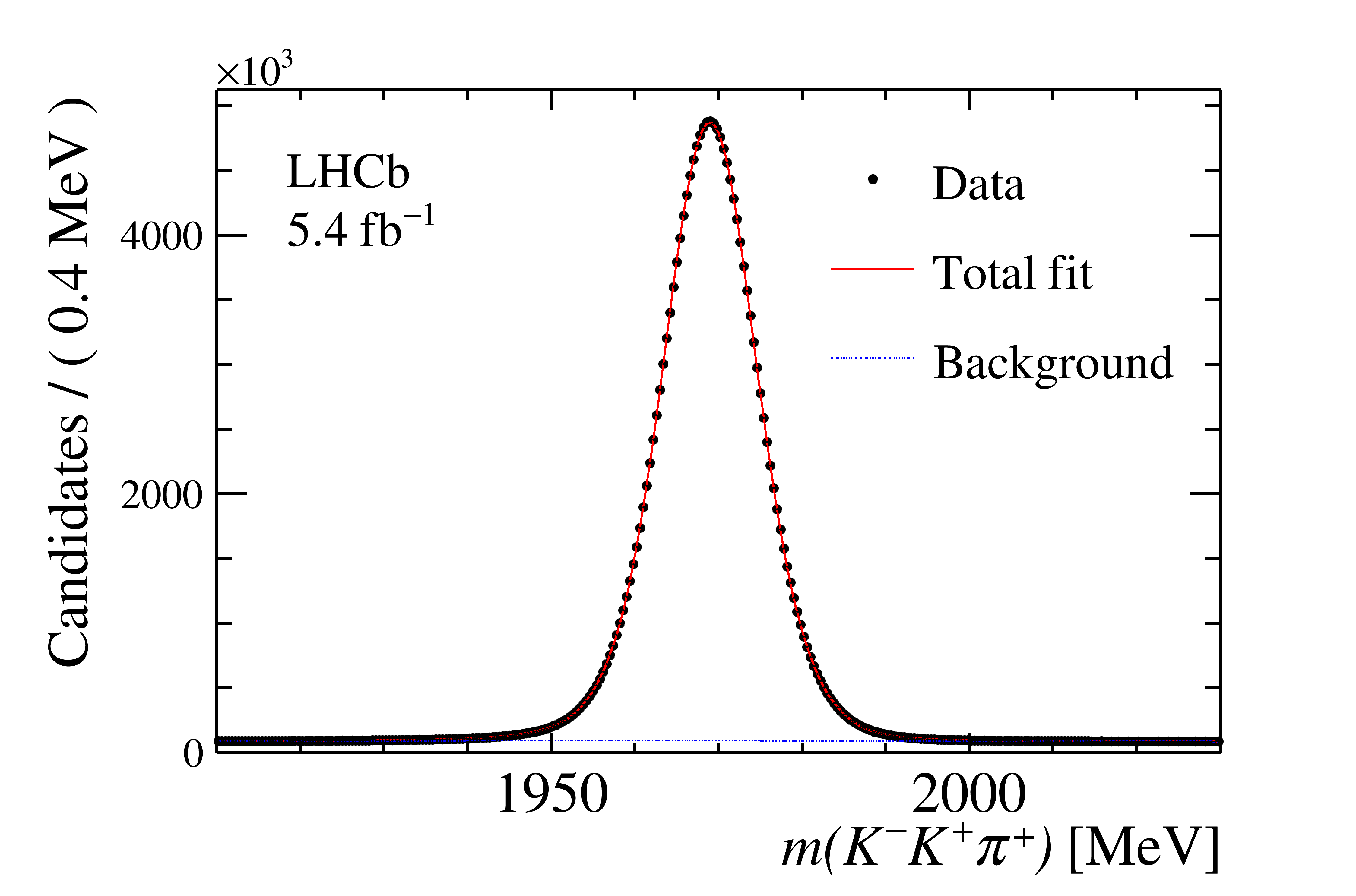} \\
        
 \end{tabular}
         \end{center}
 \caption{Invariant-mass distribution of the final samples of (left) {\ensuremath{\D^+\rightarrow K^-K^+\pi^+}\xspace} and (right) {\ensuremath{\D_s^+\rightarrow K^-K^+\pi^+}\xspace} candidates.
 }
     \label{fig:mkkpi}
 \end{figure*}

For illustration, the mass distributions of the \DptoKKP and \DstoKKP candidates, summed over all Dalitz plots bins, are given in Fig.~\ref{fig:mkkpi}.

\subsection{Uncertainties of the observables}

The statistical uncertainties of the asymmetries defined in the Letter are shown below for completeness.

\begin{equation}
  \sigma_{\Delta A^i_{\CP}} = {\sqrt{\left(1-\frac{\sigma_{\Delta A_{\rm raw}^{global}}^2}{\sigma_{ A_{\rm raw}^{\, i, S}}^2+\sigma_{ A_{\rm raw}^{\, i, C}}^2}\right)^2(\sigma_{A_{\rm raw}^{\, i, S}}^2+\sigma_{A_{\rm raw}^{\, i, C}}^2) + \sigma_{\Delta A_{\rm raw}^{global}}^4\displaystyle\sum_{j\neq i}^{N_{bins}} \frac{1}{\sigma_{ A_{\rm raw}^{\, j, S}}^2+\sigma_{ A_{\rm raw}^{\, j, C}}^2} }}
  \protect\label{eq:sigma_DeltaAi}
\end{equation}

\begin{equation}
\sigma_{\Delta A^{_{\rm raw}^{global}}} = \left[ \, \sum_i^{N_{bins}} \frac{1}{\sigma^2_{A_{\rm raw}^{\, i, S}}+\sigma^2_{A_{\rm raw}^{\, i, C}}}\, \right]^{-\frac{1}{2}}
  \protect\label{eq:sigma_DeltaA_global}
\end{equation}

%\clearpage
\subsection{Graphical representation of the \texorpdfstring{\CP}{CP} asymmetries}

\begin{figure}[bp]
\begin{center}\includegraphics[width=0.95\textwidth]{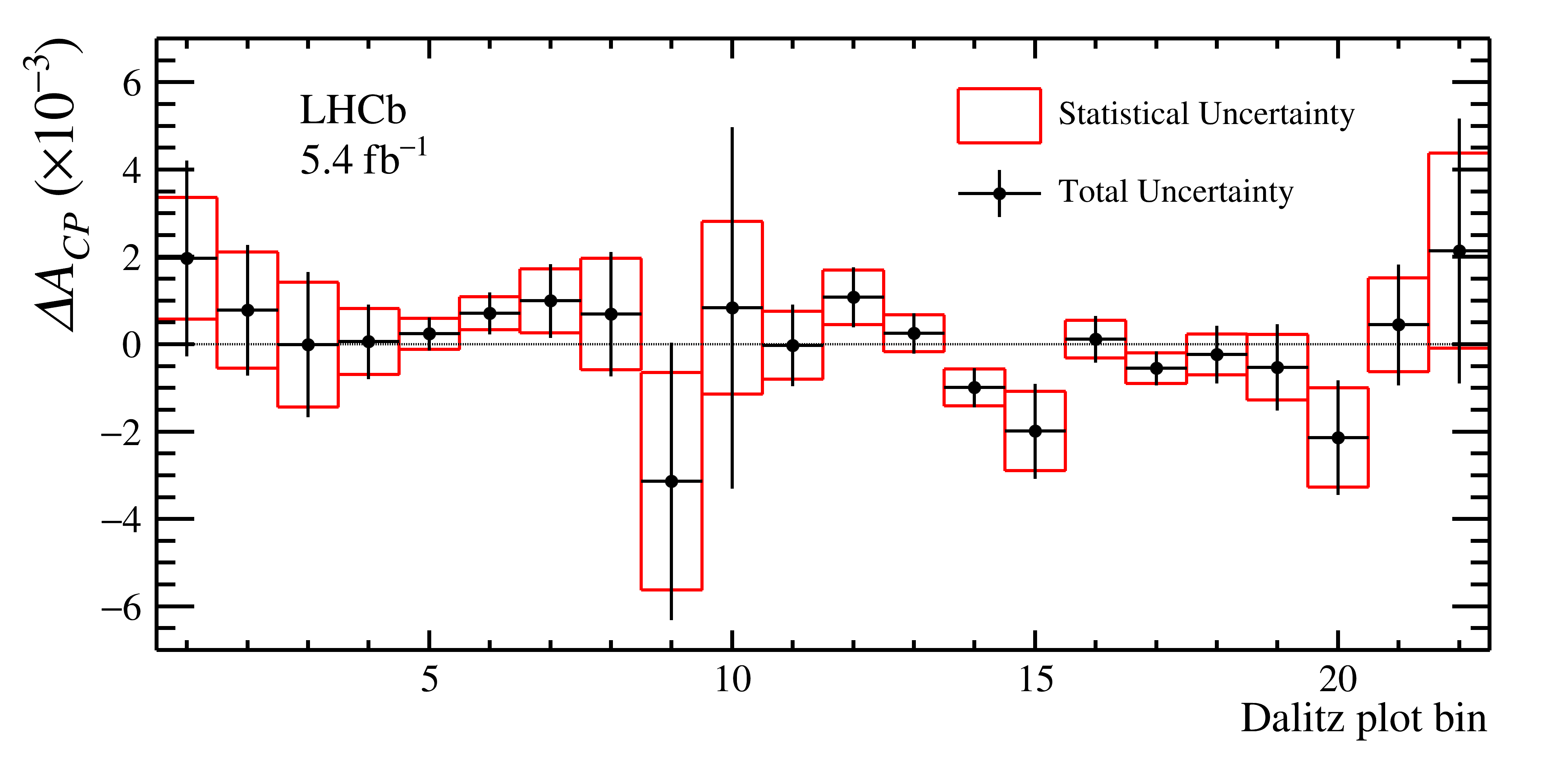}
\end{center}
    \caption{Difference of charge asymmetries ${\Delta A^i_{{{\ensuremath{C\!P}}\xspace}}}$ for each bin of the Dalitz plot.}
    \label{fig:DeltaACP}
\end{figure}

A graphical representation of the ${\Delta A^i_{\CP}}$  results is shown in Fig.~\ref{fig:DeltaACP}. The horizontal axis is given by the bin index, according to the scheme depicted in Fig.~\ref{fig:DP}.

\clearpage

\newpage
\addcontentsline{toc}{section}{References}
%\setboolean{inbibliography}{true}
\bibliographystyle{LHCb}
\bibliography{main,standard,LHCb-PAPER,LHCb-CONF,LHCb-DP,LHCb-TDR}

\newpage
% LHCb collaboration author list
% Data extracted on August 14th, 2024 at 4:05pm for paper reference LHCb-PAPER-2024-019
\centerline
{\large\bf LHCb collaboration}
\begin
{flushleft}
\small
R.~Aaij$^{36}$\lhcborcid{0000-0003-0533-1952},
A.S.W.~Abdelmotteleb$^{55}$\lhcborcid{0000-0001-7905-0542},
C.~Abellan~Beteta$^{49}$,
F.~Abudin{\'e}n$^{55}$\lhcborcid{0000-0002-6737-3528},
T.~Ackernley$^{59}$\lhcborcid{0000-0002-5951-3498},
A. A. ~Adefisoye$^{67}$\lhcborcid{0000-0003-2448-1550},
B.~Adeva$^{45}$\lhcborcid{0000-0001-9756-3712},
M.~Adinolfi$^{53}$\lhcborcid{0000-0002-1326-1264},
P.~Adlarson$^{80}$\lhcborcid{0000-0001-6280-3851},
C.~Agapopoulou$^{13}$\lhcborcid{0000-0002-2368-0147}
% ...

{\footnotesize \it

% ...
$^{13}$Universit{\'e} Paris-Saclay, CNRS/IN2P3, IJCLab, Orsay, France\\
% ...
$^{36}$Nikhef National Institute for Subatomic Physics, Amsterdam, Netherlands\\
% ...
$^{45}$Instituto Galego de F{\'\i}sica de Altas Enerx{\'\i}as (IGFAE), Universidade de Santiago de Compostela, Santiago de Compostela, Spain\\
% ...
$^{49}$Physik-Institut, Universit{\"a}t Z{\"u}rich, Z{\"u}rich, Switzerland\\
% ...
$^{53}$H.H. Wills Physics Laboratory, University of Bristol, Bristol, United Kingdom\\
% ...
$^{55}$Department of Physics, University of Warwick, Coventry, United Kingdom\\
% ...
$^{58}$School of Physics and Astronomy, University of Glasgow, Glasgow, United Kingdom\\
$^{59}$Oliver Lodge Laboratory, University of Liverpool, Liverpool, United Kingdom\\
% ...
$^{67}$Syracuse University, Syracuse, NY, United States\\
% ...
$^{80}$Department of Physics and Astronomy, Uppsala University, Uppsala, Sweden, associated to $^{58}$\\
% ...
}
\end{flushleft}

\end{document}